\documentclass[preprintnumbers,superscriptaddress,showkeys,showpacs,byrevtex]{revtex4}
\usepackage{amsmath,amsfonts,amssymb,amscd,amsxtra,amsthm}
\usepackage{graphicx}
\usepackage{bm}
\usepackage{amsmath}
\usepackage{multirow}
\newcommand {\be} {\begin{equation}}
\newcommand {\ba} {\begin{eqnarray}}
\newcommand {\ee} {\end{equation}}
\newcommand {\ea} {\end{eqnarray}}
\begin{document}
\preprint{CYCU-HEP-16-10}
\title{Meson cloud effects on the pion quark distribution function in the chiral constituent quark model}
\author{Akira Watanabe}
\email[E-mail: ]{watanabe@cycu.edu.tw}
\affiliation{Department of Physics and Center for High Energy Physics, Chung-Yuan Christian University, Chung-Li 32023, Taiwan}
\author{Chung Wen Kao}
\email[E-mail  (Corresponding Author): ]{cwkao@cycu.edu.tw}
\affiliation{Department of Physics and Center for High Energy Physics, Chung-Yuan Christian University, Chung-Li 32023, Taiwan}
\author{Katsuhiko Suzuki}
\email[E-mail: ]{katsu_s@rs.kagu.tus.ac.jp}
\affiliation{Department of Physics, Tokyo University of Science, Shinjuku, Tokyo 162-8601, Japan}
\date{\today}
\begin{abstract}
We investigate the valence quark distribution function of the pion $v^{\pi}(x,Q^2)$ in the framework of the chiral constituent quark model
and evaluate the meson cloud effects on $v^{\pi}(x,Q^2)$.
We explicitly demonstrate how the meson cloud effects affect $v^{\pi}(x,Q^2)$ in detail.
We find that the meson cloud correction causes an overall 32\% reduction of the valence quark distribution and an enhancement at the small Bjorken $x$ regime.
Besides, we also find that the dressing effect of the meson cloud will make the valence quark distribution to be softer in the large $x$ region.
\end{abstract}
\maketitle
\section{Introduction}

The parton distribution functions (PDFs) of the hadrons delineate the partonic structure of the hadrons and have been intensively studied for several decades.
However their non-perturbative nature prevents any direct application of the fundamental theory of the strong interaction, quantum chromo dynamics (QCD), since the usual perturbative technique is not applicable here.
Lattice QCD can provide some moments of the PDFs, and recently new attempts to calculate the $x$ dependencies of PDFs on the lattice have been done motivated by the X. Ji's proposal~\cite{Ji:2013dva,Chen:2016utp}, although the extraction of the light-cone PDFs from the quasi-PDFs is still under discussion.
Therefore, it remains one of the greatest challenges in hadronic physics to understand the PDFs of the hadrons
from the fundamental principles of QCD.\\

Fortunately one is still able to make connections between the hadronic phenomenology at low energy and QCD in indirect ways.
For example, the $SU(3)$ flavour symmetry of QCD provides us the clues to study the property of the hadronic bound states even though the strange quark masses
explicitly break this symmetry. Furthermore we have also learned that
the light pseudoscalar mesons are the Nambu-Goldstone bosons. It is the consequence of the spontaneously chiral symmetry breaking of the QCD vacuum.
Those Nambu-Goldstone bosons play dominant roles in the low energy hadronic structures and interactions because they will couple to particles and develop the meson clouds around the core.
The meson clouds have important impacts on the hadronic interactions and structures.\\

Actually it also explains why the constituent quark picture works so well in some aspect of
low energy hadronic phenomenology because there are always meson clouds surrounding the light quarks
such that the light quarks appear to be much heavier objects called the constituent quarks.
Furthermore, the sea-quarks and the gluons are absorbed into constituent quarks
as the components of the meson clouds. The hadrons are now treated as the bound states of these constituent quarks.
One may expect to use such a picture to
compute the PDFs of the quarks at low $Q^2$ and the sea-quarks and gluons will reappear
as the PDFs are evolved by the DGLAP equation to higher $Q^2$ in which high energy scatterings occur.
However, to apply the leading-order or even the next-to-leading-order (NLO) QCD evolution
when the initial $Q$ value is far below $1$~GeV is hard to justify because
the value of $\alpha_{s}$ is usually too large.
Hence it is necessary to take the meson cloud effects into account when one computes the PDFs in constituent quark models
since the inner structure of the constituent quarks is not completely irrelevant \cite{Melnitchouk:1994en,Kulagin:1995ia}.\\

There are many applications of this meson cloud dressing scheme to the PDFs,
but most of them are devoted to the PDFs of the nucleon.
For example, it has been found that the Goldstone boson fluctuations generate a flavour asymmetry of the quark distribution functions inside the nucleon
and significant depolarization effects reducing the fraction of the nucleon spin carried by the quarks ~\cite{Suzuki:1997wv}. However the study of the effects of Goldstone
boson fluctuation on the quark distribution functions inside the pion was less active because it is much difficult to extract the distributions experimentally. Our knowledge of the pion quark distribution is only from the Drell-Yan di-muon production by charged pion incident on the fixed nuclear target $\pi^{\pm}N\rightarrow \mu^{+}\mu^{-}X$~\cite{Conway:1989fs,Bordalo:1987cr,Betev:1985pf}.\\

Nevertheless, the pion is still one of the most important subjects in hadronic physics because of its Nambu-Goldstone character and its simplest quark content so that
there are several theoretical efforts focusing on the PDFs of the pion \cite{Shigetani:1993dx,Davidson:1994uv,Weigel:1999pc,Dorokhov:2000gu,Hecht:2000xa,
Nguyen:2011jy,Wijesooriya:2005ir,Belitsky:1996vh,Nam:2011hg,Nam:2012af,Nam:2012vm,Chen:2016sno} and some related lattice studies~\cite{Daniel:1990ah,Detmold:2003tm}.
In particular, the valence quark distribution functions of the pion have been extracted from the analysis including the next-to-leading-logarithmic threshold resummation effects in the calculation of the Drell-Yan cross section~\cite{Aicher:2010cb}. They found that at large $x$ the distribution becomes much softer than those shown in well-known PDF sets of the pion, e.g., SMRS~\cite{Sutton:1991ay} and GRS~\cite{Gluck:1999xe}. Interestingly, their results are in very good agreement with those of the preceding study based on Dyson-Schwinger equations~\cite{Hecht:2000xa}.\\

In this article we would like to investigate the meson cloud dressing effects of the quark distribution functions of the pion qualitatively.
At first we look for the bare quark distribution functions which can reproduce the phenomenological satisfactory result in Ref.~\cite{Aicher:2010cb}
after the dressing processes. Then we study how the meson cloud dressing effects modify the $x$-dependence of the bare quark distribution functions.
We also figure out how the meson cloud effects in $v^{\pi}(x,Q^2)$ change as $Q^2$ changes.\\

This article is organized as follows: We briefly review how to calculate the meson cloud dressing effects in general in Sec.~II.
In Sec.~III we explain the way we study the meson cloud effects on the pion valence quark distribution functions.
We present and discuss our numerical results in Sec.~IV and give our conclusion in Sec.~V.\\

\section{Dressing corrections to constituent quarks}

\begin{figure}[t]
\begin{tabular}{c}
\includegraphics[width=9.6cm]{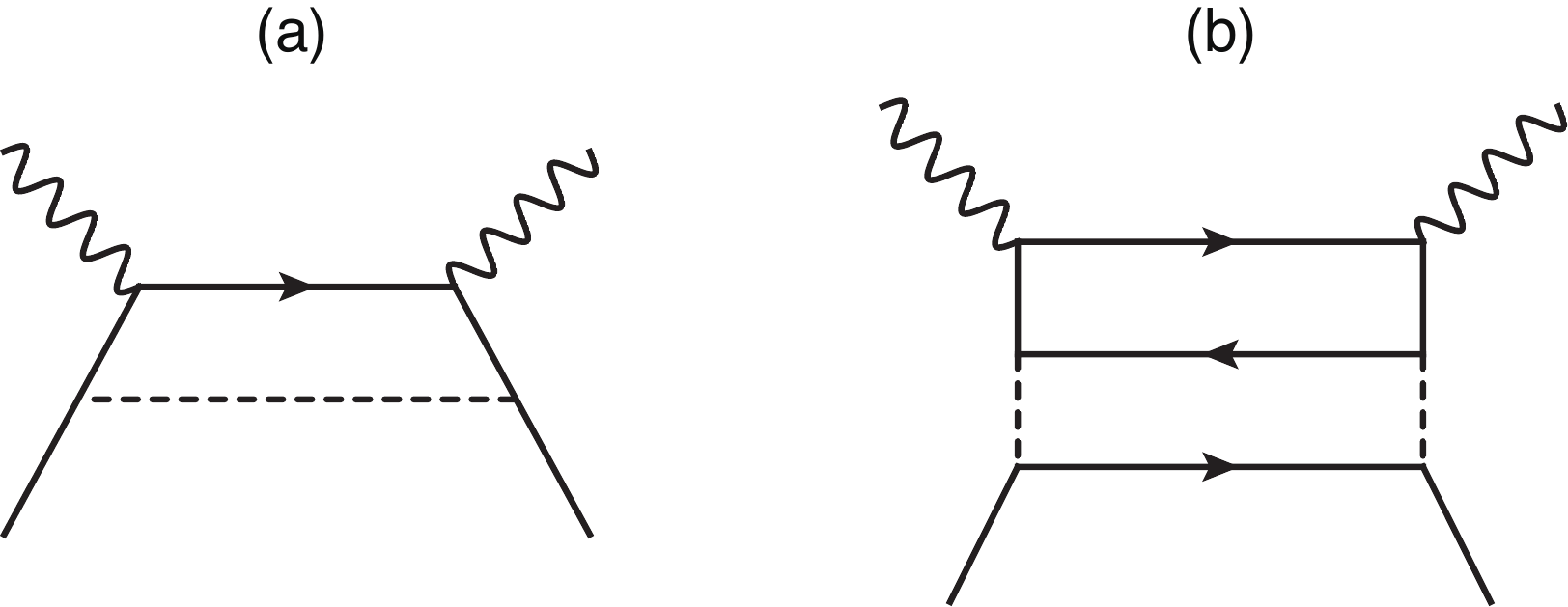}
\end{tabular}
\caption{Diagrams representing the Goldstone boson dressing corrections to a constituent quark. The wavy lines denote the
virtual photons. The thick lines denote the constituent quarks and the dashed lines represent the Goldstone bosons.
(a) describes the process in which the virtual photons couple directly to the constituent quarks and the Goldstone boson is emitted and absorbed by the
constituent quark but does not couple to the photons.
(b) describes the process where the virtual photons couple to the sea-quark pair emitted by the Goldstone boson dressing the constituent quarks.}
\label{fig:diagrams}
\end{figure}

In this section we give a brief review of the method to compute the meson cloud dressing corrections on constituent quarks following Ref.~\cite{Suzuki:1997wv}.
The interactions between the Nambu-Goldstone bosons and the constituent quarks are described by the following effective Lagrangian,
\be
{\mathcal{L}}_{int}=-\frac{g_A}{f}\bar{\psi}\gamma^{\mu}\gamma_{5}(\partial_{\mu}\Pi)\psi,
\ee
with the constituent quark field $\psi$ and the Nambu-Goldstone boson fields $\Pi$ defined as
\be
\psi=\begin{pmatrix}
u\\
d\\
s\\
\end{pmatrix}
,\,\,\,\,
\Pi=\frac{1}{\sqrt{2}}\begin{pmatrix}
\frac{\pi^{0}}{\sqrt{2}}+\frac{\eta}{\sqrt{6}} & \pi^{+} & K^{+} \\
\pi^{-} & -\frac{\pi^{0}}{\sqrt{2}}+\frac{\eta}{\sqrt{6}} & K^{0} \\
K^{-} &\bar{K}^{0} & -\frac{2\eta}{\sqrt{6}}\\
\end{pmatrix}.
\ee
Here $f$ and $g_A$ stands for the pseudoscalar decay constant and the quark axial-vector coupling constant, respectively.
The dressed and bare states of constituent quarks are related by the following relations,
\ba
|U\rangle &=& \sqrt{Z}|u_0\rangle +a_{\pi}|d\pi^{+}\rangle +\frac{a_{\pi}}{2}|u\pi^{0}\rangle +a_K|sK^{+}\rangle +\frac{a_\eta}{6}|u\eta\rangle,\nonumber \\
|D\rangle &=& \sqrt{Z}|d_0\rangle +a_{\pi}|u\pi^{-}\rangle +\frac{a_{\pi}}{2}|d\pi^{0}\rangle +a_K|sK^{0}\rangle +\frac{a_\eta}{6}|d\eta\rangle , \label{eq:dressed_states}
\ea
where $Z$ is the renormalization constant for a bare constituent quark. $|a_{\alpha}|^2$ means the probability of discovering a Goldstone boson $\alpha$ in the dressed
state of a constituent quark. The meson cloud dressing effects are depicted as the diagrams in Fig.~\ref{fig:diagrams}.
In Fig.~\ref{fig:diagrams}~(a) the Goldstone bosons around the constituent quarks are just spectators and the virtual photons directly couple to the constituent quarks.
In Fig.~\ref{fig:diagrams}~(b) the virtual photons couple to the sea-quark pair emitted by the Goldstone boson dressing the constituent quarks.
Those diagrams are evaluated in the infinite momentum frame so that the calculation will be simplified a lot by neglecting the Z-graph terms.
The quark distribution with the correction corresponding to Fig.~\ref{fig:diagrams}~(a) is given by
\be
{q_j}\left( x \right) = \int_x^1 {\frac{{dy}}{y}} {P_{j\alpha /i}}\left( y \right){q_i}\left( {\frac{x}{y}} \right),
\label{eq:1}
\ee
where $P_{j\alpha /i}$ is the splitting function which gives the probability to find a constituent quark $j$ carrying the momentum fraction $y$ together with a spectator Goldstone boson $\alpha$, whose parent constituent quark is denoted by $i$.
With the masses of the constituent quarks and the pseudoscalar mesons, $m_i$, $m_j$, and $m_{\alpha}$, $P_{j\alpha /i}$ is expressed as
\be
{P_{j\alpha /i}}\left( y \right) = \frac{1}{{8{\pi ^2}}}{\left( {\frac{{{g_A}\bar m}}{f}} \right)^2}\int {dk_T^2} \frac{{{{\left( {{m_j} - {m_i}y} \right)}^2} + k_T^2}}{{{y^2}\left( {1 - y} \right){{\left( {m_i^2 - M_{j\alpha }^2} \right)}^2}}}.
\label{eq:Pjalphaoveri}
\ee
$M_{j\alpha }^2$ is the invariant mass squared of the final state expressed  as
\be
M_{j\alpha }^2 = \frac{{m_j^2 + k_T^2}}{y} + \frac{{m_\alpha ^2 + k_T^2}}{{1 - y}},
\ee
and ${\bar m} = (m_i + m_j )/2$ is the average of the constituent quark masses.
To perform the integration of Eq.~\eqref{eq:Pjalphaoveri}, a cutoff is required. Therefore
we replace $g_A$ by the following function~\cite{Speth:1996pz},
\be
{g_A}\exp \left( {\frac{{m_i^2 - M_{j\alpha }^2}}{{4{\Lambda ^2}}}} \right),
\ee
where $\Lambda$ is the cutoff parameter.
The moments of the splitting functions are defined by
\be
\left\langle {{x^{n - 1}}{P_{j\alpha /i}}} \right\rangle  \equiv \int_0^1 {dx} {x^{n - 1}}{P_{j\alpha /i}}\left( x \right),
\ee
and the first moments with $n=1$,
\be
\left\langle {{P_{j\alpha /i}}} \right\rangle  = \left\langle {{P_{\alpha j/i}}} \right\rangle  \equiv \left\langle {{P_\alpha }} \right\rangle ,
\ee
are also defined.
The renormalization constant $Z$ is then expressed as
\be
Z = 1 - \frac{3}{2}\left\langle {{P_\pi }} \right\rangle  - \left\langle {{P_K}} \right\rangle . \label{eq:renormalization_constant}
\ee
Here only the pion and kaon contributions are considered.
Since the contribution from $\eta$ is relatively small compared to them due to the factor $1/6$ which appears in Eqs.~\eqref{eq:dressed_states}, we simply neglect it in this study.

The diagram Fig.~\ref{fig:diagrams}~(b) represents the process in which the sea-quark pair emitted from Goldstone boson is probed. The
corresponding contribution is given by
\be
{q_k}\left( x \right) = \int {\frac{{d{y_1}}}{{{y_1}}}\frac{{d{y_2}}}{{{y_2}}}} {V_{k/\alpha }}\left( {\frac{x}{{{y_1}}}} \right){P_{\alpha j/i}}\left( {\frac{{{y_1}}}{{{y_2}}}} \right){q_i}\left( {{y_2}} \right).
\label{eq:2}
\ee
Here $P_{\alpha j/i} (x) = P_{j \alpha /i} (1 - x)$ and $V_{k/\alpha }$ is the quark distribution function with the flavor index $k$ in the Goldstone boson $\alpha$, for which the normalization condition,
\be
\int_0^1 {dx} {V_{k/\alpha }}\left( x \right) = 1,
\ee
is satisfied.

\section{Dressing corrections to the valence quark distribution functions of pion}

In this section we will apply the procedure in the previous section to the pion valence quark distribution functions.
To evaluate the meson cloud dressing corrections, we choose the values for some parameters shown in Table~\ref{table:parameters}.
\begin{table}[tb]
\caption{The set of parameters chosen in this article.}
\begin{center}
\begin{tabular}{| c | c | c | c | c | c | c | c |} \hline
$g_A$ & $m_u$ & $m_d$ & $m_s$ & $m_\pi$ & $m_K$ & $f_\pi$ & $\Lambda$ \\
\hline \hline
1.0 & 360 MeV & 360 MeV & 570 MeV & 140 MeV & 494 MeV & 93 MeV & 1.4 GeV \\
\hline
\end{tabular}
\end{center}
\label{table:parameters}
\end{table}
These are typical values guided by preceding studies based on the Nambu--Jona-Lasinio model~\cite{Nambu:1961tp,Nambu:1961fr,Hatsuda:1994pi}.
It has been observed for the nucleon that the obtained value of the Gottfried sum rule with this parameter set in the present framework is consistent with the experimentally observed empirical one~\cite{Suzuki:1997wv}.

\begin{figure}[tb!]
\begin{center}
\includegraphics[width=0.7\textwidth]{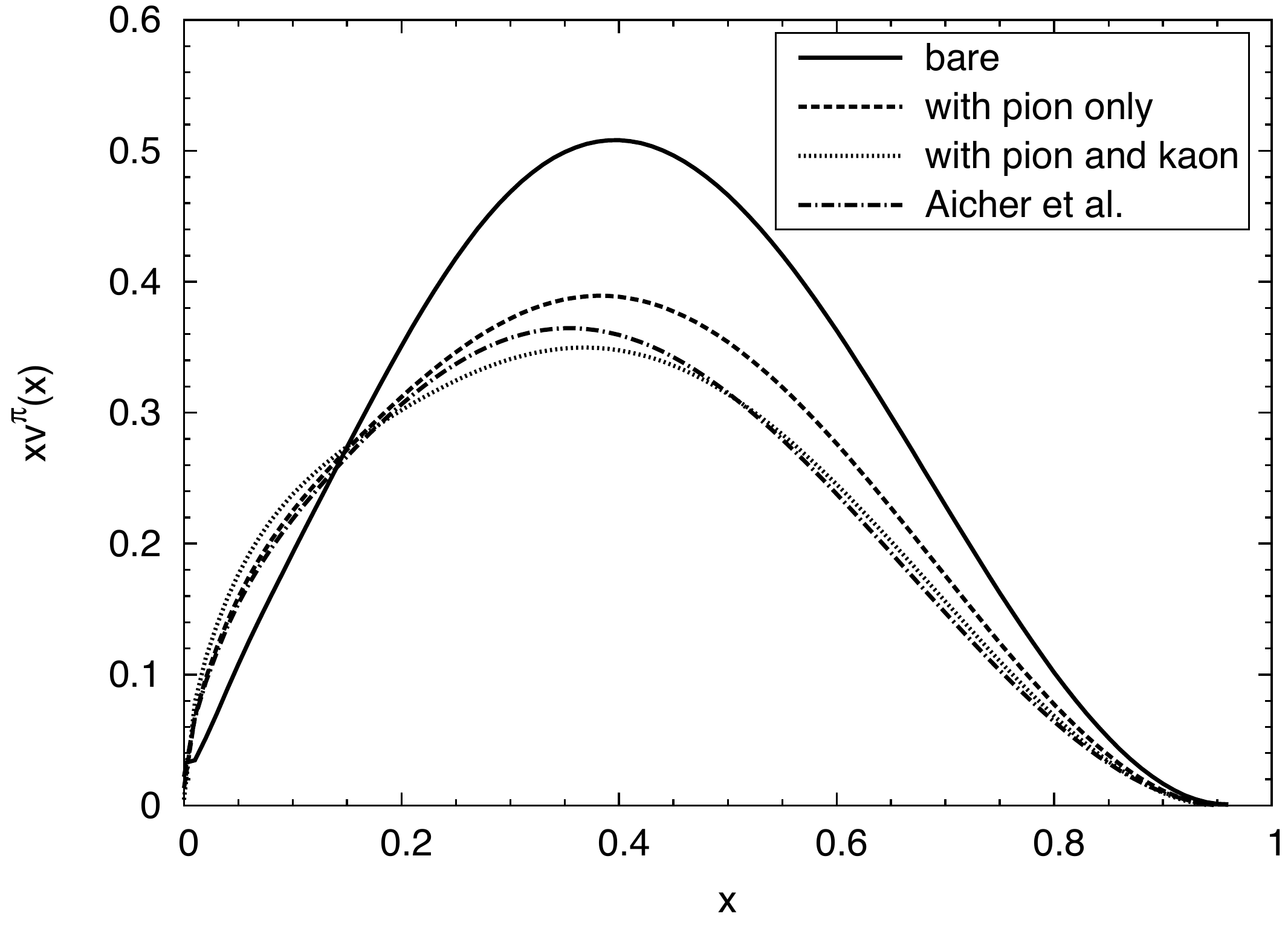}
\caption{The valence quark distribution functions of the pion at $Q = 4$~GeV as a function of Bjorken $x$.
The solid curve shows the bare distribution $v^{\pi}_{\rm bare}(x)$.
The dashed and dotted lines represent the dressed $v^{\pi}(x)$ including only the pion dressing effect, and both the pion and kaon dressing effects, respectively.
The dashed-dotted line represents the empirical curve provided in Ref.~\cite{Aicher:2010cb}.
}
\label{fig:piplus_original}
\end{center}
\end{figure}
Our first task is to choose the appropriate bare quark distribution functions of $\pi ^+$, $u_0 (x)$ and $\bar d _0 (x)$ satisfying the normalization conditions,
\be
\int_0^1 {dx} {u_0}\left( x \right) = \int_0^1 {dx} {\bar d_0}\left( x \right) = 1.
\ee
Through this article we
adopt the following short hand notation,
\be
P \otimes q \equiv \int_x^1 {\frac{{dy}}{y}} P\left( y \right)q\left( {\frac{x}{y}} \right).
\ee
According to Eq.~\eqref{eq:1} and Eq. \eqref{eq:2}, after the dressing the distribution functions of $u$ and $\bar u$ quarks in $\pi ^+$ become as follows,
\begin{align}
u\left( x \right) = &Z{u_0}\left( x \right) + \frac{1}{2}{P_{u\pi /u}} \otimes {u_0} + {V_{u/\pi }} \otimes {P_{\pi d/u}} \otimes {u_0} + {V_{u/\pi }} \otimes {P_{\pi \bar u/\bar d}} \otimes {{\bar d}_0} \nonumber \\
&+ \frac{1}{4}{V_{u/\pi }} \otimes {P_{\pi u/u}} \otimes {u_0} + \frac{1}{4}{V_{u/\pi }} \otimes {P_{\pi \bar d/\bar d}} \otimes {{\bar d}_0} + {V_{u/K}} \otimes {P_{Ks/u}} \otimes {u_0}, \\
\bar u\left( x \right) = &{P_{\bar u\pi /\bar d}} \otimes {{\bar d}_0} + \frac{1}{4}{V_{\bar u/\pi }} \otimes {P_{\pi u/u}} \otimes {u_0} + \frac{1}{4}{V_{\bar u/\pi }} \otimes {P_{\pi \bar d/\bar d}} \otimes {{\bar d}_0}.
\end{align}
We then define the valence quark PDF of the pion as
$v^{\pi}_{\rm dressed}(x) = u^{\pi^{+}}_{v}(x) = \bar{d}^{\pi^{+}}_{v}(x) = d^{\pi^{-}}_{v}(x) = \bar{u}^{\pi^{-}}_{v}(x)$,
and obtain the following expression,
\ba
v^{\pi}_{\rm dressed}\left( x \right)&=&u\left( x \right) - \bar u\left( x \right) \nonumber \\
&=&Z{u_0}\left( x \right) + \frac{1}{2}{P_{u\pi /u}} \otimes {u_0} + {V_{u/\pi }} \otimes {P_{\pi d/u}} \otimes {u_0} + {V_{u/\pi }} \otimes {P_{\pi \bar u/\bar d}} \otimes {{\bar d}_0} \nonumber \\
&+& {V_{u/K}} \otimes {P_{Ks/u}} \otimes {u_0} - {P_{\bar u\pi /\bar d}} \otimes {{\bar d}_0}.
\label{eq:pion}
\ea
Integrating both sides of Eq.~\eqref{eq:pion}, we obtain
\be
\int_0^1 {dx} {v^{\pi}_{\rm dressed}} \left( x \right) = \int_0^1 {dx} \left\{ {u\left( x \right) - \bar u\left( x \right)} \right\}
= Z + \frac{3}{2}\left\langle {{P_\pi }} \right\rangle  + \left\langle {{P_K}} \right\rangle
= 1,
\ee
which shows the correct normalization, following Eq.~\eqref{eq:renormalization_constant}.\\

To make the physics to be more transparent, we divide the dressed valence PDF into four terms according to their origins,
\be
v^{\pi}(x,Q^2)=v^{\pi,0}(x,Q^2)+v^{\pi,a}(x,Q^2)+v^{\pi,b1}(x,Q^2)+v^{\pi,b2}(x,Q^2),
\label{eq:eachterms}
\ee
where each term is defined as follows,
\ba
v^{\pi,0}(x,Q^2)&=&Z{u_0}\left( x \right), \nonumber \\
v^{\pi,a}(x,Q^2)&=&\frac{1}{2}{P_{u\pi /u}} \otimes {u_0}- {P_{\bar u\pi /\bar d}} \otimes {{\bar d}_0},\nonumber \\
v^{\pi,b1}(x,Q^2)&=&{V_{u/\pi }} \otimes {P_{\pi d/u}} \otimes {u_0} + {V_{u/\pi }} \otimes {P_{\pi \bar u/\bar d}} \otimes {{\bar d}_0}, \nonumber \\
v^{\pi,b2}(x,Q^2)&=&{V_{u/K}} \otimes {P_{Ks/u}} \otimes {u_0}.
\label{eq:eachterms2}
\ea
$v^{\pi,a}(x,Q^2)$ is the contribution from Fig.~1~(a).
$v^{\pi,b1}(x,Q^2)$ and $v^{\pi,b2}(x,Q^2)$ are the contributions from Fig.~1~(b) with the pion cloud and Fig.~1~(b) with the kaon cloud, respectively.
For simplicity, we assume here the same initial $u$ quark distribution of the involved $K ^+$ as pion's.
To perform the $Q^2$ evolution, we utilize the code produced by Kobayashi, Konuma, and Kumano~\cite{Kobayashi:1994hy}, and the results presented in the next section
are obtained at NLO accuracy.

\begin{figure}[tb!]
\begin{center}
\includegraphics[width=0.7\textwidth]{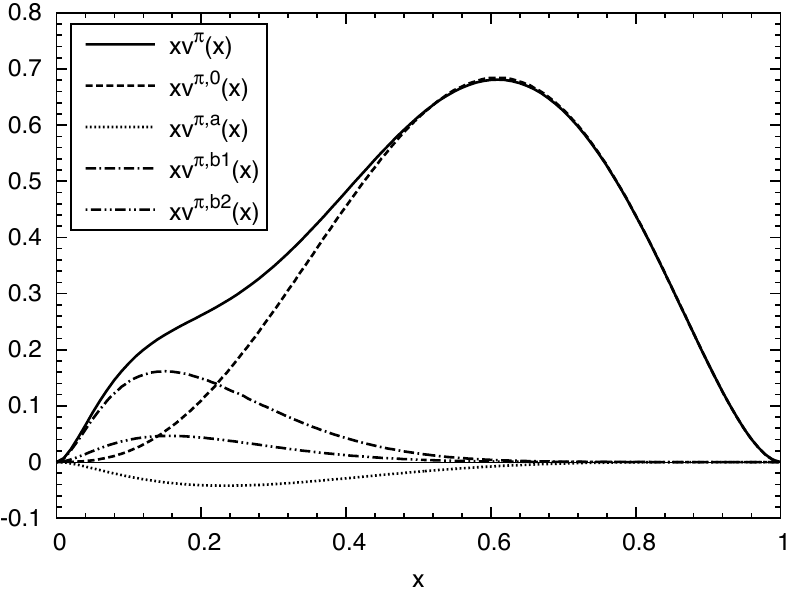}
\caption{The contributions of each term in Eq.~\eqref{eq:eachterms} to $xv^{\pi}_{\rm dressed}(x)$ at $Q = Q_0$.
The dashed, dotted, dashed-dotted and dashed-double dotted lines represent the contributions of the first term $v^{\pi,0}(x)$, second term $v^{\pi,a}(x)$, third term $v^{\pi,b1}(x)$ and fourth term $v^{\pi,b2}(x)$, respectively.
The solid line represents the sum of the each terms.}
\label{fig:eachterms}
\end{center}
\end{figure}

\section{Results and Discussion}
In this section we present and discuss our numerical results.
First of all, we shall fix the bare valence quark distribution function of the pion $v^\pi_{\rm bare} (x,Q_0^2)$.
We assume that only the valence quarks exist at the initial scale, and start with the symmetric distribution, $v^\pi_{\rm bare} (x,Q_0^2) \propto x^\alpha (1-x)^\alpha$.
We try to find the value of $\alpha$ by fitting $v^\pi_{\rm dressed} (x,Q^2)$ to the distribution obtained from the fit 3 in Ref.~\cite{Aicher:2010cb} at $Q = 4$~GeV, varying the initial scale $Q_0$.
Since our purpose of this study is to discuss the meson cloud effects qualitatively, we simply choose an integer value for $\alpha$, and then fix the bare distribution as
\be
v^\pi_{\rm bare} (x,Q_0^2) = N x^{1.8} (1 - x)^{1.8},
\ee
where $N$ is the normalization constant.
The initial scale $Q_0=0.50$~GeV is determined by fitting the resulting first moment to
that of Aicher et al. at $Q = 4$~GeV~\cite{Aicher:2010cb}.
Our bare $v^{\pi}(x,Q^2)$ at $Q=4$~GeV is depicted as the solid line in Fig.~\ref{fig:piplus_original}.
We also present the dressed $v^{\pi}(x,Q^2)$ with only pion cloud (in dashed line)
and the one fully dressed with the pion and kaon clouds (in dotted line) at $Q=4$~GeV.
One can compare our full result
with the phenomenologically satisfactory one (in dashed-dotted line) given by Ref.~\cite{Aicher:2010cb}.
Our full result agrees with the empirical one excellently
when $x \ge 0.5$ and becomes a slightly smaller between $x=0.4$ and 0.8, and slightly lager when $x\le 0.2$. However the difference between our result and the empirical curve
is always less than few percents such that our study in this article is qualitatively reliable and not just a prototype study.
\begin{figure}[tb!]
\begin{center}
\includegraphics[width=0.7\textwidth]{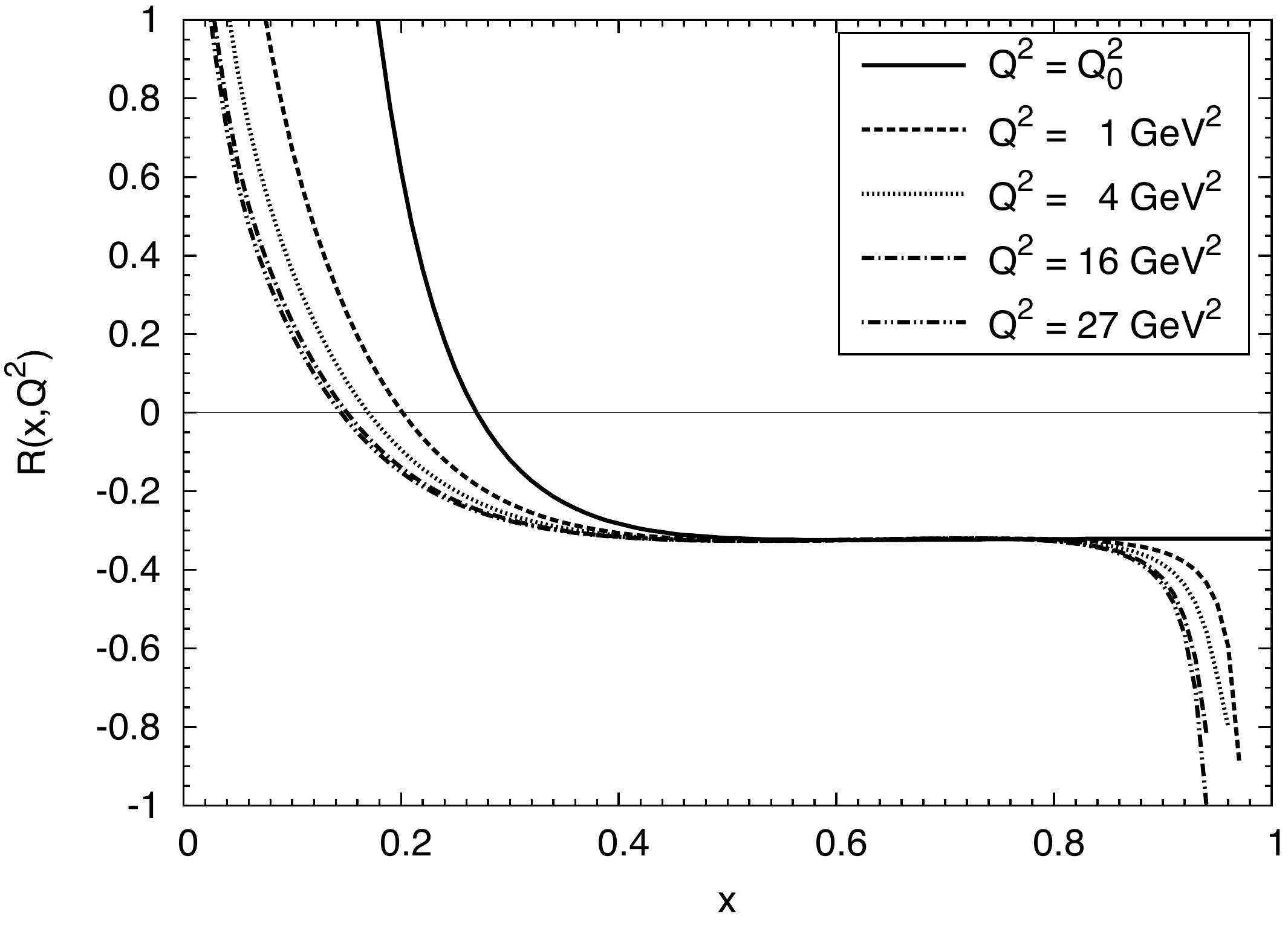}
\caption{
The ratio of the valence quark distribution functions of the pion, $R(x,Q^2)$ defined in Eq.~\eqref{eq:R},
as a function of Bjorken $x$ for various $Q^2$.
The solid, dashed, dotted, dashed-dotted and dashed-double dotted lines stand for $Q^2 = Q_0^2$, 1, 4, 16, 27~GeV$^2$ cases, respectively.
}
\label{fig:correction_strength_from_original}
\end{center}
\end{figure}

Fig.~\ref{fig:eachterms} provides a clear anatomy of the meson cloud dressing effects on the pion valence quark distribution function
at $Q = Q_0$ since the contribution of each term in  Eq.~\eqref{eq:eachterms} is demonstrated explicitly there.
The solid curve corresponds to $xv^{\pi,0}(x)$ which is just the bare valence quark PDF multiplying the renormalization factor
$Z$. The value of $Z$ is about $2/3$. Actually one finds that $xv^{\pi,0}(x)$ and the total dressed $xv^{\pi}(x)$ are almost identical when $x \ge 0.5$.
Namely, at $Q = Q_0$ the meson cloud dressing effect is exclusively the effect of the renormalization factor $Z$ in the large $x$ regime.
The contribution from Fig.~1~(a) represented in the dotted line in Fig.~\ref{fig:eachterms} is negative and its magnitude is relatively small compared with $xv^{\pi,0}(x)$
and $xv^{\pi,b1}(x)$.
On the contrary, the contribution from Fig.~1~(b) involving the pion cloud
is positive and much larger than $xv^{\pi,a}(x)$ in the magnitude. It is represented by the dashed-dotted line in Fig.~\ref{fig:eachterms}.
But it is only nonzero within $0 \le x \le 0.6$. Moreover its effect is cancelled by the $v^{\pi,a}$ in the regime $0.4\le x\le 0.6$.
Hence its effect is most significant in the regime $0\le x\le 0.4$.
The contribution of Fig.~1~(b) involving the kaon cloud represented in the dashed-double dotted line in Fig.~\ref{fig:eachterms}
is small but positive. Its contribution
is largely cancelled by the $xv^{\pi,a}(x)$ between $x = 0$ and $0.2$.
Hence the curve corresponding to the total pion valence quark distribution function at $Q = Q_0$ is just the sum of
$xv^{\pi,0}(x)$ and $xv^{\pi,b1}(x)$ approximately.\\

Naively one may think that there is no dressing effect in the large $x$ regime except the overall 32\% reduction according to the above analysis.
Nevertheless the further analysis shows otherwise. To observe the total dressing effect one is able to define the following quantity,
\be
R(x, Q^2) = \frac{v^{\pi}_{\rm dressed} (x,Q^2) - v^{\pi}_{\rm bare} (x,Q^2)}{v^{\pi}_{\rm bare} (x,Q^2)}.
\label{eq:R}
\ee
The quantity characterizes the extent of the modification of the valence quark distribution function due to the Goldstone boson dressing effects.
Our results are demonstrated in Fig.~\ref{fig:correction_strength_from_original}.
At $Q=Q_0$ indeed the dressing effect is simply overall 32\% reduction due to the renormalization factor $Z$ in the large $x$ region.
However, when one goes to a higher $Q$ value, a suppression of $R(x,Q^2)$ appears in the region of $x\ge 0.8$.
Namely, the curves of $R(x,Q^2)$ drop rather dramatically at the right end of the Fig.~\ref{fig:correction_strength_from_original}.
This suppression becomes more pronounced as the $Q^2$ increases.
Since the large $x$ behavior of the valence quark PDF usually can be characterized by some exponent $\beta$ such that $v^{\pi}(x)\sim (1-x)^{\beta}$ as $x\sim 1$. The dressing effect will increase the value of $\beta$ substantially as $Q^2$ increases.
Furthermore, there is a plateau between $x = 0.4$ and $0.8$ where the dressing effect is just overall 32\% reduction for all curves at different $Q^2$ values.
On the other hand, in the small $x$ regime, the Goldstone boson dressing effect is to enhance the $v^{\pi}(x)$ significantly.
But the magnitude of this enhancement decreases very fast as $Q^2$ increases.
For example, one can find in Fig.~\ref{fig:correction_strength_from_original}, at $x=0.2$ and $Q = 1$~GeV the enhancement of $v^{\pi}$
is about 30\%, but the enhancement is reduced to only 15\% at $Q=2$~GeV, moreover, at $Q=4$~GeV the enhancement becomes only 10\%.
The enhancement is mainly due to the contribution of the Fig.~1~(b) involving the pion cloud according to Fig.~\ref{fig:eachterms}.
We conclude that the meson cloud dressing effect on the valence quark distribution function is an overall 32\% reduction with an enhancement at the small $x$ regime $0\le x\le 0.4$ and an additional suppression in the $x\ge 0.8 $ region.
This generic feature is shared for the curves with different $Q^2$ values.

\section{Conclusion}
In this article, we have investigated the meson cloud dressing effects on the pion valence quark distribution function $v^{\pi}(x,Q^2)$ in detail.
We first choose the bare $v^{\pi}(x,Q^2)$ judiciously such that the dressed $v^{\pi}(x,Q^2)$ agrees with the phenomenologically satisfactory $v^{\pi}(x,Q^2)$
given in Ref.~\cite{Aicher:2010cb}, then we study the individual contribution of the meson cloud dressing effects. We find that the meson cloud effect is mainly
an overall 32\% reduction due to the renormalization factor $Z$ with a suppression in the large $x$ regime, $x\ge 0.8$, and an enhancement in the small $x$ region,
$x\le 0.4$. The extent of the enhancement decreases fast as $Q^2$ increases, on the contrary, the magnitude of the suppression increases as $Q^2$ increases.
We also find that the enhancement at the small $x$ regime is mainly from the pion cloud contribution.\\

Currently, COMPASS Collaboration at CERN is taking data of the pion-induced Drell-Yan process.
Absolute production cross sections will be measured there, and the preceding predictions for the valence quark distribution function of the pion will be tested with the data. Besides, the pion-induced Drell-Yan experiment is also planned at J-PARC in Japan. In this experimental facility, the beam energy is much lower compared to that of the COMPASS experiment, so the cross sections are sensitive to the larger $x$ region. Hence, our understanding of the valence quark distribution of the pion in the region will be improved so that our understanding of the bare quark distribution function also will be improved via the framework presented in this article.
\section*{Acknowledgments}
We would like to thank Hsiang-nan Li and Wen-Chen Chang for their valuable suggestions and comments.
A.W. is supported by the grant MOST 105-2811-M-033-004 of Ministry of Science and Technology, Taiwan.
C.W.K. is supported by the grants NSC 102-2112-M-033-005-MY3 and MOST 105-2112-M-033-004 of Ministry of Science and Technology, Taiwan.


\end{document}